\documentclass[twocolumn,showpacs,amsfonts]{revtex4-1}

\usepackage{amsmath}
\usepackage{latexsym}
\usepackage{float}
 \usepackage{amssymb}
\usepackage{graphicx}
\usepackage{textcomp}
\usepackage{hyperref}
\usepackage{subfigure}
\usepackage{color}

\def\ba{\begin{eqnarray}}
\def\ea{\end{eqnarray}}
\def\be{\begin{equation}}
\def\ee{\end{equation}}
\def\bm{\begin{math}}
\def\me{\end{math}}

\newcommand{\dummy}

\begin{document}
\title{Dependence of Cluster Growth on Coefficient of Restitution  in a Cooling Granular Fluid}
\author{Subir K. Das$^{1,2,*}$ and Subhajit Paul$^{1,3}$}
\affiliation{$^1$ Theoretical Sciences Unit, Jawaharlal Nehru Centre for Advanced Scientific Research,
 Jakkur P.O, Bangalore 560064, India}
\affiliation{$^2$ School of Advanced Materials, Jawaharlal Nehru Centre for Advanced Scientific Research,
 Jakkur P.O, Bangalore 560064, India}
\affiliation{$^3$ Institut f\"ur Theoretische Physik, Universit\"at Leipzig, Postfach 100920, D-04009, Leipzig, Germany}

\date{\today}

\vskip 20cm
\par
\begin{abstract}
Starting from configurations having homogeneous spatial density, we
study kinetics in a two-dimensional system of inelastically colliding hard particles, 
a popular model for cooling granular matter. Following an initial time period, 
the system exhibits a crossover to an inhomogeneous regime that is characterized by the formation
and growth of particle-rich clusters. We present
results on the time dependence of 
average mass of the clusters and that of average kinetic energy, obtained via event driven molecular dynamics simulations, 
for a wide range of values for the coefficient of restitution ($e$), by fixing the overall density of particles 
in the system to a constant number. 
	The time of onset of crossover from homogeneous to the inhomogeneous regime, as is well known, strongly increases as one 
moves towards the elastic limit. Nevertheless, our presented results suggest that the asymptotic 
growth is independent of $e$, for uniform definition of cluster, onset of which has a different $e$-dependence 
than the onset of above mentioned crossover. In other words, not only the exponent but also the amplitude
of the power-law growth, which is widely believed to be the form of the evolution, is at the most very weakly
sensitive to the choice of $e$. While it is 
tempting to attribute this fact to the similar feature in the decay of energy, we caution that our
current understanding is not matured enough to draw such a connection between cluster
growth and  energy decay in a meaningful manner.
\end{abstract}



\maketitle
\section{ Introduction}

~~Granular materials \cite{aranson, bril} consist of particles of varying shapes and sizes, and are 
very commonly observed in nature. Typical examples \cite{aranson,bril,bri1,lmat} are powders, drugs, sacks of rice or sugar grains, 
packets of coffee beans, cosmic dust, etc. Thus, knowledge of the behavior of granular materials is of immense 
importance in many disciplines. Understanding of these, at different 
length and time scales, can be useful in the interpretation of the formation of planetary rings; has 
applications in industries like pharmaceutical, agriculture and mining; is of importance in prevention 
of damages due to natural processes like landslide, erosion, etc. This, however, is
challenging, a reason being that  often these materials share properties of both
fluids and solids \cite{aranson, bril}. 
\par
~~Due to friction and inelastic collisions among particles these 
systems continuously cool, i.e., particles loose kinetic energy, average value of which defines the granular 
temperature. This leads to interesting pattern formation that, for a class of systems 
\cite{aranson, bril, gold, nie, ben, luding1, brito, haff, bodrova, das1, das2, shinde1, paul1, paul2, mcnamara, camp, luding2, her1,
tak1, che1, mil1}, 
resembles \cite{bray, binder1, roy, majumder} coexistence of particle-rich and particle-poor clusters
during vapor-liquid transitions. Over the past few 
decades there have been intense research activities to identify and understand the form of energy decay and 
cluster growth in this class of systems. Focus of the present paper is related to these. 
\par 
~~In this context, in the original form of a popular model, to be referred to as the granular gas model (GGM) \cite{gold}, 
energy dissipation occurs only due to inelastic collisions among hard constituents,
the coefficient of restitution ($e$) lying between $0$ and $1$. This dissipation leads to 
progressive parallelization of velocities of the particles and formation of clusters in the so called 
inhomogeneous cooling state (ICS). Like in kinetics of phase transitions \cite{bray, binder1}, here also 
typically one asks: How does the average mass ($m$) of these clusters grow with time ($t$)?  
There is a reasonably fair belief that the growth is of power-law type
\cite{luding1, das1, das2, paul1, paul2}:
\begin{equation}\label{mass}
 m \sim t^{\zeta}.
\end{equation}
In the ICS, the decay of average kinetic energy ($E$) 
is even more widely studied aspect. It has been observed that 
this also follows power-law \cite{nie, ben, shinde1, majumder}: 
\begin{equation}\label{energy}
E \sim t^{-\theta}.
\end{equation}
\par 
~~There has been immense interest in estimation of and understanding the 
dependence of $\zeta$ and $\theta$ on space dimension ($d$) and other system parameters like particle
density ($\rho$) and $e$. In this paper, $\rho$ is calculated as $N/V$, where
$N$ is the number of particles and $V$ is the volume of the box.
Furthermore, establishing connection 
between $\theta$ and $\zeta$ also remains of significant current research interest.
With respect to this, while good progress has been made in $d=1$, the status is much inferior for 
higher dimensions. In $d=1$ there exists evidence that \cite{nie, ben, shinde2, carnevale, paul3}
\begin{equation}\label{1d_exp}
m \sim 1/E \sim t^{2/3},
\end{equation}
irrespective of the values of $\rho$ and $e$. In fact, GGM in this dimension is believed to be equivalent to 
another popular model (perhaps simpler, though extremely useful), referred to as the ballistic aggregation model (BAM) \cite{carnevale}. 
\par 
~~In the BAM hard spherical particles move ballistically and following a collision the partners 
merge to form a larger spherical entity, keeping the mass and momentum conserved. For this 
model Carnevale {\it{et al.}} \cite{carnevale}  predicted that 
\begin{equation}\label{bam_exp}
m \sim 1/E \sim t^{2d/(d+2)},
\end{equation}
implying strong, inverse relation between clustering and dissipation in all dimensions.
Computer simulations, however, reported discrepancies \cite{paul3, trizac1, trizac2} with this 
prediction in $d>1$, when packing fraction is not too high. For the BAM another theory predicts 
that \cite{trizac1, trizac2}
\begin{equation}\label{hyper}
 2\zeta + d\theta =2d.
\end{equation}
At least up to $d=3$, it has been observed in simulations with different packing fractions that this 
hyperscaling relation is valid \cite{paul2}. 
\par 
~~While no such strong and accurate connection between cluster growth 
and energy decay for the GGM has been established, undoubtedly  the 
decay of the latter is the cause for the growth of the former. In this work we present results 
from the simulation study of this model in $d=2$ for a wide range of $e$. Our results on  
the growth of mass are suggestive of certain interesting universal feature. 
Similar feature is observed in the decay of $E$ as well. 
Nevertheless, we are reluctant to draw connection between the two. We cite example (from $d=1$) to 
emphasize that the relation between energy decay and growth of mass is rather  
complex in GGM. Thus, understanding of the observation requires further attention.
\par 
~~The rest of the paper is organized as follows. In section II we provide further details of the model and 
describe certain methods. The results are presented in section III. Finally, section IV concludes the paper with a 
brief summary and outlook.

\section{Model and Methods}
\begin{figure}[htb]
\centering
\includegraphics*[width=0.4\textwidth]{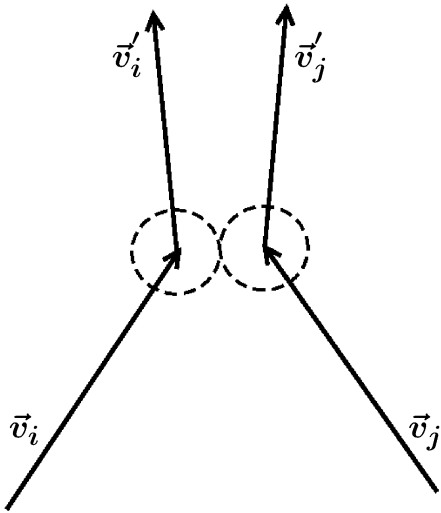}
\caption{\label{fig1}  
Sketch of a collision event for $e<1$ in $d=2$. The particles ($i$ and $j$)
are represented by dashed circles. The arrow-headed lines represent velocities of the 
particles before ($\vec{v}_i$, $\vec{v}_j$) 
and after ($\vec{v}_i^{'}$, $\vec{v}_j^{'}$)
the collision. This picture is only for the purpose of qualitative demonstration and drawn
without reference to a coordinate system.
}
\end{figure}

~~In the standard two-dimensional GGM \cite{gold}, that we consider here, a system consists of equisized inelastic 
hard discs.  The velocities of colliding partners $i$ and $j$ before and after (represented with prime) 
an instantaneous collision are related via the equations \cite{allen}
\begin{equation}\label{ivelo}
 \vec{v}^{'}_{i} = \vec{v}_{i} - \frac{1+e}{2} [\hat{n}\cdot(\vec{v}_{i}-\vec{v}_{j})]\hat{n},
\end{equation}
and
\begin{equation}\label{jvelo}
 \vec{v}^{'}_{j} = \vec{v}_{j} + \frac{1+e}{2} [\hat{n}\cdot(\vec{v}_{i}-\vec{v}_{j})]\hat{n}.
\end{equation}
Here $\hat{n}$ is an unit vector aligned with the relative position of the colliding partners.
Equations (\ref{ivelo}) and (\ref{jvelo}) satisfy the conservation of momentum and
contain the fact that there is collisional energy dissipation by a factor $1-e^2$.
For $e<1$, velocities of the partners become more parallel after
a collision. This is qualitatively depicted in Fig. \ref{fig1}.
\par 
~~With this rule, we have performed event-driven molecular dynamics simulations 
\cite{allen, rapa}. 
After every new collision, the task of the simulation code is to identify the  
partners and instant for the next collision.
Between collisions these particles move ballistically, i.e., with constant velocities. 
Progress of time is calculated by adding the intervals between collisions \cite{nie, ben}. 
While this method provides the real time, in
the literature dynamics of this model has been quantified by using this time as well as 
with respect to the number of collisions per particle \cite{das1, das2}. We do not adopt 
the latter here. Even though there exists linear relation \cite{nie, paul1} between these two 
times in the ICS, this is not the case during the homogeneous period. 
Such discrepancy or nonuniformity between the two regimes, with respect to the connection
between two different measures of time,
occurs due to the following fact. At early regime,
compared to the late time situation, the systems contain mostly faster moving particles.
The velocity distributions \cite{bodrova} are different in the two regimes  
with large regions having velocities of particles aligned with each
other at late time.

\par 
~~At late time one encounters serious technical problem with this simulation method, particularly 
for low values of $e$. Often collisions remain restricted to a tiny group of 
neighboring particles with small relative velocities. This fact, referred to as the inelastic collapse \cite{mcnamara}, severely 
limits the progress of time. A method \cite{ben, mcnamara, camp, luding2} to overcome this problem 
considers assignment of $e=1$ for collisions corresponding to relative speed smaller than a cut-off value $\delta$. 
For $d>1$, however, this problem is less severe. So, most  of our results were obtained
by employing $\delta=0$. 
\par 
~~All our simulations started with random initial configurations in both position and velocity, with 
Maxwellian distribution for the latter \cite{allen, rapa}. For each set of parameter values the
starting temperature was same. We have applied periodic boundary conditions 
in all possible directions. The quantitative results are presented after averaging over at least $10$ independent initial configurations.
\par 
~~The clusters were identified as regions having density over a certain critical number \cite{paul1, paul2},
chosen to be same for all values of $e$.
Boundaries around the clusters were appropriately marked to facilitate the calculation of number of 
particles within a cluster as well as the estimation of the corresponding mean value which is the average mass ($m$). 
The results for the energy correspond to the average kinetic energy, calculation of which is straight-forward.

\section{ Results}

\begin{figure}[htb]
\centering
\includegraphics*[width=0.4\textwidth]{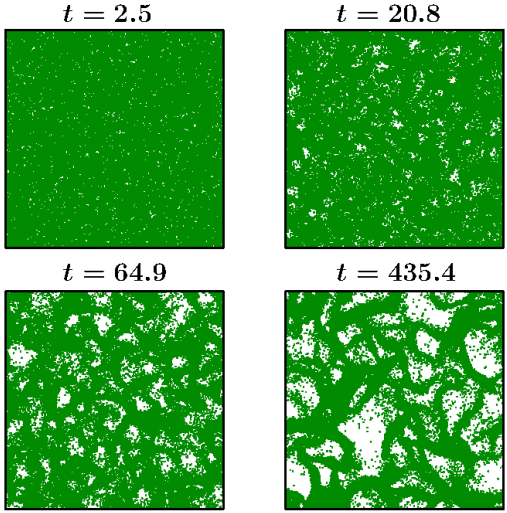}
\caption{\label{fig2} Snapshots, taken during the evolution of the freely cooling granular gas in space 
dimension $d=2$, are presented for coefficient of restitution $e=0.8$ and overall particle density $\rho=0.37$. 
The location of the particles are marked. At the top of each of the frames corresponding time is 
mentioned.}
\end{figure}
~~In Fig. \ref{fig2} we present evolution snapshots for the considered model in $d=2$. Frames from four different 
times of a particular run are shown. The results correspond to $\rho=0.37$ and $e=0.8$. For equisized discs 
of diameter unity, this value of $\rho$ corresponds to a packing fraction of approximately $0.29$.
For the earliest presented time, 
i.e., at $t=2.5$, the particles are still homogeneously distributed over the entire system. By $t=20.8$ 
crossover to the ICS has started. Regions rich and poor in particles are clearly identifiable 
from the snapshot at $t=64.9$, average size of which has grown significantly in the last snapshot.
\begin{figure}[htb]
\centering
\includegraphics*[width=0.42\textwidth]{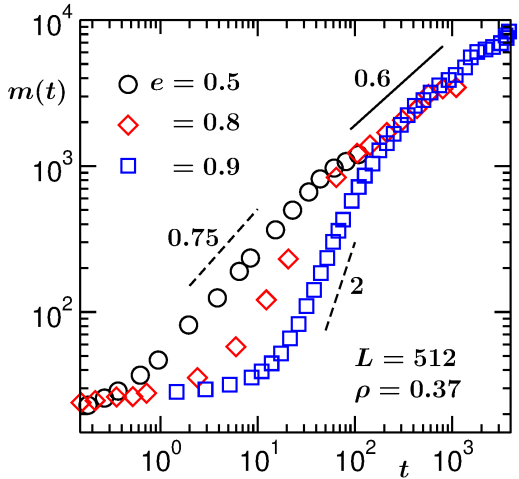}
\caption{\label{fig3} The average mass of clusters ($m$) is plotted versus time, on a log-log 
scale, for overall particle density $\rho=0.37$. Results from multiple values of $e$ are 
	included. The solid and dashed lines represent power-laws. The values of the exponent are 
mentioned in appropriate places.}
\end{figure}
\par 
~~This growth is quantitatively depicted in Fig. \ref{fig3}. Here we have plotted average mass as a function 
of time, on a log-log scale. Data from three different values of $e$, covering a rather wide range, 
have been presented. In each of the cases, we have $\rho=0.37$. Given that the linear dimension
($L$, in units of the particle diameter) of our square simulation box is $512$, the results correspond to $N=96993$.
Since energy dissipation through 
inelastic collisions is the reason behind the clustering phenomena \cite{gold}, it is expected \cite{das1, das2} 
that the onset of ICS will occur earlier for smaller values of $e$. This fact can easily be appreciated 
from the displayed set of results. With the increase of $e$, values of $m$ remains stable
at a small number, that corresponds to random, homogeneous distribution of particles, 
over longer periods of time.
\par 
~~At late time the reasonable linear appearance of the data sets on the log-log scale hints towards power-law behavior. 
The consistency of the data with the solid line suggests that \cite{paul2}
\begin{equation}\label{mass_exp}
\zeta \simeq 0.6.
\end{equation}
There has been longstanding interest in the community in estimating the exponent for this growth. 
Few other works \cite{paul2, carnevale, paul3}, combined with these results, point to the possibility that the value of 
the exponent is `practically' independent of $d$, $\rho$ and $e$. This contradicts both Eqs. (\ref{bam_exp}) 
and (\ref{hyper}). Here note that  various authors \cite{nie, paul2} showed that the energy decay 
for the present model follows Eq. (\ref{bam_exp}), implying
\begin{equation}\label{ke_exp}
	\theta=1,
\end{equation}
in $d=2$. Thus, the decay of $E$ and the growth of $m$ are not generally connected to each other via Eq. (\ref{bam_exp}), 
clearly stating the nonequivalence between GGM and BAM in $d=2$ (and dimensions higher than that). 
The hyperscaling relation of Eq. (\ref{hyper}), 
in a fixed dimension, has its relevance with respect to the density dependence of the two exponents. 
This relation, or anything analogous, also does not appear 
to be true for the GGM when results from other studies in $d>1$ are looked at \cite{paul2}.
\par 
~~While the above results and discussions are mostly related to strengthening  of certain previously observed facts, 
the new interesting observation of the present study is the following.  
The data sets in Fig. \ref{fig3} appear to overlap with each other at long times. 
This hints towards the fact that the scaling growths for all
the $e$ values are same, not only in the exponent $\zeta$ but also in the amplitude. We repeat, onset of
the crossover to the ICS gets delayed with the increase of $e$. E.g., for $e=0.5$ the onset 
occurs at a time less than unity ($t\simeq 0.35$), whereas for $e=0.9$ the crossover starts at $t\simeq3$.
Nevertheless, all the data sets overlap at late time and the overlapping times appear disproportionate
to those for the onset of crossover. Note that the ratio of the two times corresponding to the onset
of crossover for largest and the smallest $e$ values is approximately $9$.
On the other hand, the ratio of the times when these data sets start showing consistency with the $t^{0.6}$ behavior
is approximately $3.5$. This is due to sharper growth, during the crossover period, 
for larger $e$ value. The latter point can be clearly appreciated from  
Fig. \ref{fig3}. For the presented range of $e$ values the (approximate) power-law exponent in this regime
changes from $0.75$ (for $e=0.5$) to $2$ (for $e=0.9$). This is an interesting fact in itself.
\par
At this point it will 
be useful to investigate the structural aspect in the asymptotic regime for different values of $e$. Outcome of this 
may lead to a more unique statement about the growth, involving mass as well as morphology.

\begin{figure}[htb]
\centering
\includegraphics*[width=0.42\textwidth]{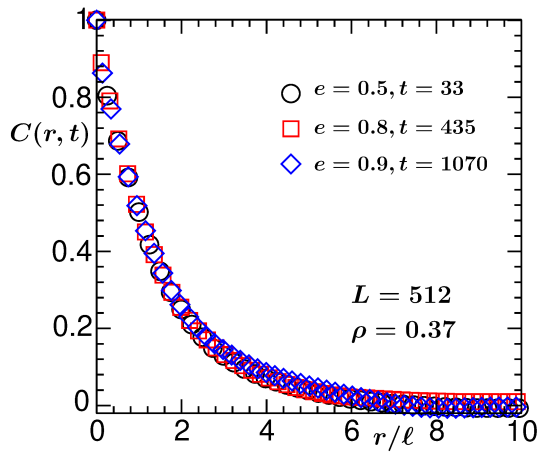}
\caption{\label{fig4} Normalized two-point equal time correlation function, $C(r,t)$, is plotted versus the 
scaled distance $r/\ell$, $\ell$ being the average linear dimension of the clusters. We have shown 
data from three different values of $e$. In each of the cases times are chosen in such a way 
	that the systems are in the scaling regime of growth (see Fig. \ref{fig3}). All data sets are for $\rho=0.37$.}
\end{figure}

\par 
~~In Fig. \ref{fig4} we show plots of (normalized) two-point equal time correlation function \cite{bray}, $C(r,t)=C_o(r,t)/C_o(0,t)$, with 
\begin{equation}\label{correl}
C_o(r,t)= \langle\psi(\vec{r},t) \psi(\vec{0},t)\rangle - \langle\psi(\vec{r},t)\rangle\langle\psi(\vec{0},t)\rangle,
\end{equation}
as a function of $r/\ell(t)$, where $r=|\vec{r}|$ is the scalar distance between two space points 
and $\ell(t)$ is the average linear dimension of the clusters at a given time $t$. In 
Eq. (\ref{correl}), $\psi$ is an appropriate order parameter \cite{bray, fisher1, golden}, values of which 
identify the particle-poor and particle-rich regions. This we have defined as 
\begin{equation}\label{opm}
	\psi (\vec{r},t) = \mbox{sgn}(\rho_{\rm loc}(\vec{r},t) - \rho_c),
\end{equation}
where $\rho_{\rm loc}(\vec{r},t)$ is the local particle density at a space point $\vec{r}$ at time $t$ and 
$\rho_c$ is a cut-off density which we have set \cite{paul1, paul2} to $0.5$. The behavior of 
$C(r,t)$ provides information on the character of a structure and is commonly used in the literature 
of phase transitions \cite{bray, binder1, fisher1, golden}.
\par 
~~The average linear dimension, $\ell$, of the structure can be estimated from the decay 
of $C(r,t)$, say, as 
\begin{equation}\label{crl_cross}
	C(r=\ell,t)=a,
\end{equation}
where $a$ is a pre-assigned constant, having a value less than $1$. In this work, however, we have estimated $\ell$ via a 
different route, viz., by exploiting the domain size distribution function, $P(\ell_d,t)$, as \cite{roy, majumder}
\begin{equation}\label{dmn_sze}
 \ell = \int \ell_d P(\ell_d,t) d\ell_d,
\end{equation}
where $\ell_d$ is the distance between two successive interfaces (between low and high density regions) 
along any Cartesian direction. There exist other methods as well in the literature \cite{bray}. Each of them provides value 
differing by only constant factors from the others. One needs, in this regard, to be careful that for 
comparative purposes unique method must be adopted.
\par 
~~In Fig. \ref{fig4} we have included results from all three values of $e$. 
In each of the cases the times are 
chosen from the long-time power-law regime. The collapse of data, upon rescaling the distance axis by $\ell$, 
confirms that the structure is also similar [at least in a coarse-grained, hard-spin sense that is
embedded in the calculation of $\rho_{\rm loc}$ and definition of $\psi$ in Eq. (\ref{opm})]
for all values of $e$ in the asymptotic growth regime.
Analogous results were presented in an earlier work \cite{das1, das2} on both growth and structure. 
However, in that work the range of $e$ was narrower and naturally the robustness of the 
phenomena, combining, on one hand, vastly different crossover times and on the other,
unique asymptotic growth, could not have been captured. Also, in that work the `real' time was not used 
for the quantification of growth of `length'.

\begin{figure}[htb]
\centering
\includegraphics*[width=0.42\textwidth]{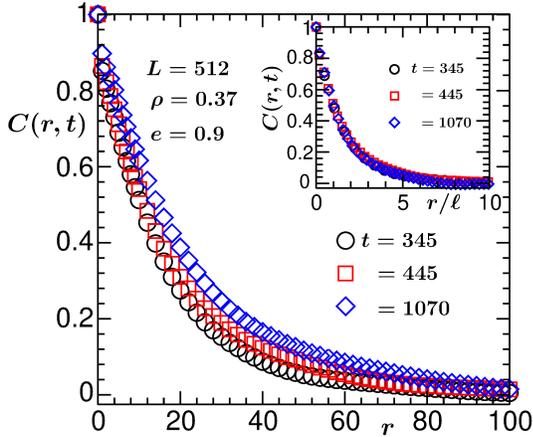}
\caption{\label{fig5} Plots of $C(r,t)$ are shown versus $r$, for $e=0.9$ and $\rho=0.37$. Data from
three different times are included. The inset shows same data sets but here the
distance axis is scaled by $\ell$.}
\end{figure}

For completeness, next we demonstrate that there exists self-similarity in structure with time,
a standard practice in studies of coarsening phenomena.
For that one requires to realize superposition of data for $C(r,t)$ from different times
when plotted versus $r/\ell$. 
In Fig. \ref{fig5} we show a representative set of results,
for $e=0.9$ and $\rho=0.37$. In the main frame we show direct plots, i.e.,  $C(r,t)$ versus $r$,
from three different times. Clearly, with increasing time the decay is getting slower,
implying growth in the system. In the inset we have demonstrated nice overlap of data
from all the three times
by scaling the distance axis by $\ell$. This confirms self-similar growth
in the power-law regime of Fig. \ref{fig3} for $e=0.9$. The same is true for other
values of $e$ as well. However, for brevity we do not present those results.

\begin{figure}[htb]
\centering
\includegraphics*[width=0.42\textwidth]{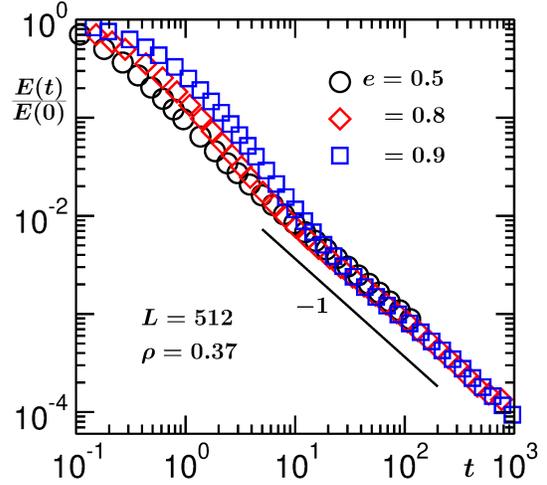}
\caption{\label{fig6} Log-log plots of average kinetic energy versus time are seen. Data from 
three different values of $e$, for $\rho=0.37$, are shown. The solid line is a power-law. 
The value of the exponent is mentioned next to it.}
\end{figure}
\par 
~~For possible explanation of this unique evolution, in Fig. \ref{fig6} we present log-log plots of kinetic 
energy versus time. Again results for all three values of $e$ are shown. The late time behavior, i.e., 
energy decay in the ICS, is consistent with $E \sim t^{-1}$, for each of the $e$ values, which is in agreement 
with Eq. (\ref{bam_exp}) or Eq. (\ref{ke_exp}), that was also observed by other authors \cite{nie, paul2}. 
The form of the decay prior to this is different and 
referred to as the Haff's cooling law \cite{haff}. 
It is identifiable from this 
figure as well that with increasing $e$ appearance of ICS gets delayed.

\begin{figure}[htb]
\centering
\includegraphics*[width=0.42\textwidth]{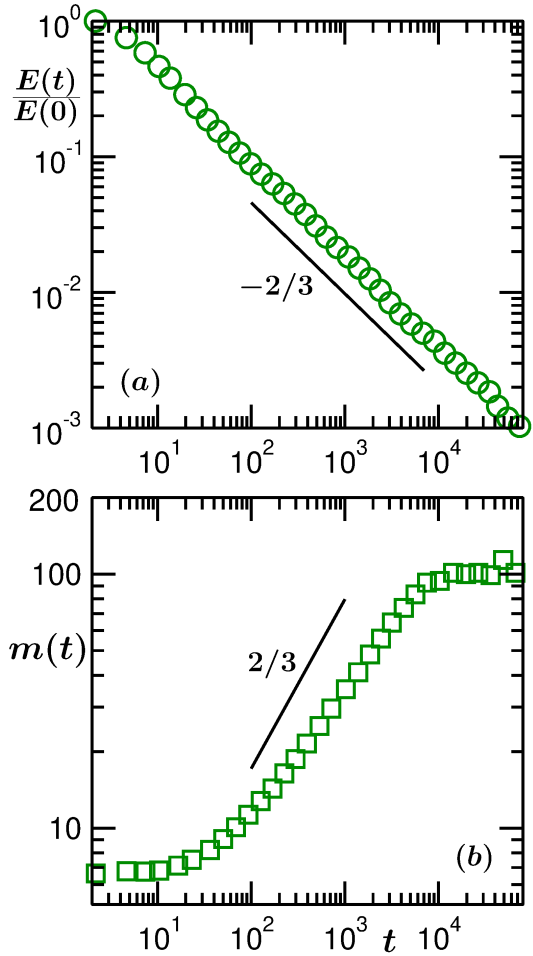}
\caption{\label{fig7} Log-log plots of (a) average kinetic energy and (b)
average cluster mass, versus time, for the GGM in $d=1$.
All results are for $e=0.5$, $\rho=0.3$ and $\delta=0.005$. The
solid lines represent power-laws. The exponents are mentioned next to the respective lines.
The results are similar to those in Fig.1 of Phys. Rev. E {\bf 96}, 012105 (2017).}
\end{figure}
\par 
~~Here also ICS data from different $e$ values superimpose on top of each other \cite{nie}.
(In the Haff's regime, for different $e$ values the deviations
from each other is expected.)
Even though we have discussed that $E$ and $m$ cannot be connected via Eq. (\ref{bam_exp}) or 
Eq. (\ref{hyper}), decay of the former is the reason for growth of the latter,
in the ICS, beyond doubt. Thus, one may 
argue that overlapping feature of mass  can be explained via that in the energy data. To counter 
this, we would like to discuss an example from $d=1$ (cf. Ref. \cite{shinde2, paul3}) 
to emphasize the fact that a connection between decay of energy 
and growth of mass is quite puzzling for GGM. The intention is to stress upon the fact
that the explanation of the universal feature described above is not straight-forward.
\par 
~~In Fig. \ref{fig7} we present results from $d=1$: Part (a) contains data for 
decay of energy and part (b) shows the growth of mass.  
Results for both energy and mass are for fixed 
density and coefficient of restitution (see caption for these numbers). It appears that the growth 
of mass has frozen while the energy decay continues for much longer 
with the same exponent, that is consistent with the theoretical 
expectation of Eq. (\ref{1d_exp}) or Eq. (\ref{bam_exp}).
Even though we have used a nonzero $\delta$ here, these results 
nevertheless demonstrate the presence of immense complexity in this simple  
model of granular matter.

In Fig. \ref{fig7} even though the energy decay is clearly seen to be consistent
with $\theta=2/3$, for the growth of mass the value of $\zeta$ does not appear to
be $2/3$. However, via advanced methods of analysis, including a renormalization
group technique, it was confirmed \cite{paul3} that $\zeta\simeq 2/3$.

\section{Conclusion}

~~From the event-driven molecular dynamics \cite{allen, rapa} simulations we have presented results on the kinetics 
in a granular gas model \cite{gold}. In this model energy dissipation and velocity parallelization 
occur due to inelastic collisions among constituent particles. This leads to clustering phenomena, resembling 
the kinetics in a chemical system undergoing vapor-liquid transition \cite{majumder}. The onset of clustering strongly depends upon 
\cite{das1,das2} the overall 
particle density ($\rho$) in the system and coefficient of restitution ($e$).
\par 
~~The key result of this paper is related to the dependence of growth of mass, in the long time limit, on 
the coefficient of restitution. Strikingly, for a fixed overall density we observe that despite strong 
dependence of the onset of clustering on $e$, the asymptotic growth is same, i.e., if the
character is of power-law the values of growth exponent and 
amplitude appear to be similar for all the presented values of $e$ that cover a rather wide range.
\par
~~Of course, better statistics and more accurate analysis are 
necessary to put our conclusion on a concrete footing. Nevertheless, even in its current form this observation requires attention.
A possible route for explanation of the phenomena could be 
the similar observation in the decay of energy. But we argue by providing example that the connection 
between energy decay and cluster growth in GGM may be more complex than realized.
\par
~~Studies \cite{hummel} analogous to this were performed in granular gases via direct numerical simulations of
Navier-Stokes equation. The authors of this work looked at the universality in density fluctuations in
the clustering phenomena with respect to the variations of different model parameters. 
It will be interesting to check this for GGM as well for different $e$ values. This
will be useful in understanding whether the universal feature that we report here for mass should
also hold for characteristic length. 
\par
~~More such studies are necessary
to characterize universal features in granular materials. E.g., as an extension of the present work, 
we intend to explore a spectrum of $\rho$ and $e$ in different dimensions, in future. 
Even though it is not expected that growth data from different densities will collapse with each other 
in the inhomogeneous cooling regime, it will be interesting to check for the relevant scaling factors to 
obtain a master curve. 

$^*$ das@jncasr.ac.in

\end{document}